\documentclass[aps,prb,twocolumn,showpacs]{revtex4-1}

\usepackage{graphicx}
\DeclareGraphicsExtensions{.png}

\usepackage{xcolor}
\usepackage{amsmath}

\usepackage{dcolumn}
\usepackage{bm}
\usepackage{hyperref}

\begin{document}

\title{Quantum spin Hall phase in multilayer graphene}

\author{N. A. Garc\'ia-Mart\'inez}
\author{J. L. Lado}
\author{J. Fern\'andez-Rossier} 
\altaffiliation{On leave from Departamento de F\'isica Aplicada, Universidad de Alicante,  Spain}
\affiliation{International Iberian Nanotechnology Laboratory (INL),
Av. Mestre Jos\'e Veiga, 4715-330 Braga, Portugal
}

\date{\today} 

\begin{abstract}
The so called quantum spin Hall phase is a topologically non trivial
insulating phase that is predicted to appear in graphene and
graphene-like systems. In this work we address the question of
whether this topological property persists in multilayered
systems. We consider two situations:  purely multilayer graphene
and heterostructures where graphene is encapsulated by trivial
insulators with a strong spin-orbit coupling.
We use a four orbital tight-binding model that includes the
full atomic spin-orbit coupling and we 
calculate the $Z_{2}$ topological invariant of the bulk states
as well as the edge states of semi-infinite crystals with armchair termination.
For homogeneous multilayers we find that even when the spin-orbit
interaction opens a gap for all the possible stackings, only those
with odd number of layers host gapless edge states while those with even
number of layers are trivial insulators.
For the heterostructures where graphene is encapsulated by trivial
insulators, it turns out that the interlayer coupling is able to
induce a topological gap whose size is controlled by the spin-orbit
coupling of the encapsulating materials, indicating that the quantum spin
Hall phase can be induced by proximity to trivial insulators.
\end{abstract}

\maketitle

\section{Introduction}
In their seminal papers,\cite{Kane2005} Kane and Mele established the
existence of two fundamentally different types of band insulators with
time reversal symmetry in two
dimensions, dubbed as trivial and topological.
Remarkably, it was predicted that monolayer graphene
would be topological, giving rise to protected
chiral gapless edge states. 
Importantly, this opened a new venue in condensed
matter physics, the quest of searching and designing topological states
in two dimensional  systems. 

The nature of the topological state in graphene
comes from the
intrinsic spin-orbit coupling (SOC). In particular,
SOC will
open gaps of opposite signs at the two Dirac points, in contrast
with the trivial gap that a staggered potential  opens
in the honeycomb lattice,  with the same sign at the two valleys.
This twisting of the wave functions in the reciprocal space
leads to the appearance of in-gap states at the boundaries of the
material. Subsequent work\cite{Min2006,Yao2007,Gmitra2009} found
that the size of the SOC gap in graphene was very small, and
the attention shifted to other systems, such as CdTe/HgTe
quantum wells\cite{Bernevig2006a}
in which the quantum spin Hall (QSH) phase was
found\cite{Konig2007} as well as to bulk systems, for which the
notion of topologically non-trivial insulators was extended. Experimental
evidence for  quantum spin Hall phase has also been found in
other systems, such as Bi(111) atomically
thin layers\cite{Sabater2013,Drozdov2014}, and InSb/GaSb quantum
wells\cite{Zhang2014,Knez2011}.

Multilayers of two dimensional materials are also potential candidates
to sustain topological states. 
In particular, their appealing comes from the
the tunability of stacking a different number of layers,
or even different materials.
In the present work we will focus on the study of
a particular type of multilayer systems,
whose basic building blocks
are graphene-like systems.
We will study mainly two families of multilayers.
First we consider multilayered systems formed by graphene-like
insulators using the SOC 
as a free parameter, so the main concepts should be suitable for
systems such as graphene,  Silicene\cite{Vogt2012,Fleurence2012,Xu2012},
Germanene\cite{Houssa2010} or Stanene\cite{Xu2013} (in fact
our methods make it easy to extended this kind of analysis to
the case of Bismuth\cite{Murakami2006,Sabater2013,Drozdov2014}, and
metal-organic
frameworks\cite{Zhou2014,Davila2014,Lalmi2010,Liu2011,Tang2014,Petit2009,Rodenas2014}).
Second, stacks formed by a layer of graphene encapsulated by some
trivial insulator with a strong SOC.

From a practical point of view, 
several reasons motivate this work. First, there is a generic
interest in the possibility of engineering the electronic
properties of two dimensional crystals, such as graphene, h-BN
and transition metal dichalcogenides, by combining them into
multilayers\cite{Giovannetti2007,Kosmider2013,Geim2013}.
Stacking monolayers of the same type is also a very interesting
and widely studied possibility.

Our second motivation is to study the behavior of the topological
gap as we increase the number of layers in the system.
In the case of graphene, it is well
known that key electronic properties, such as the pattern of Landau
levels and the density of states at the Dirac point are drastically
modified for bilayer\cite{McCann2013} and trilayer
graphene.\cite{McCann2006,Konschuh2011a,Zhang2010,Coletti2013,Min2008}.
Recent experimental work shows that some sort of magnetic order can
occur, even at $B=0$ in bilayer\cite{Freitag2012,Velasco2014},
trilayer\cite{Lee2014} and even tetralayers\cite{Grushina2015}.  
These last trivial symmetry breaking states, will compete with
the potential topological states studied in our work.

A third motivation comes from recent
experiments\cite{Avsar2014} that report a very large enhancement
of the spin Hall effect for graphene deposited on top of WS$_2$, as
trivial semiconductor with a quite large SOC.  This
inspires the calculation for graphene placed between two insulators
with a trivial band gap, large SOC and broken
inversion symmetry, to mimic the properties of WS$_2$ and related
transition metal dichalcogenides.

Furthermore, we have also a formal motivation. It is not obvious a priori that
the original second-neighbor hopping Hamiltonian
\cite{Kane2005}
can be  applied to multilayer graphene\cite{Cortijo2010}. In monolayer
graphene the $p_z$ orbitals are strictly decoupled from the $s$, $p_x$, $p_y$
orbitals, due to mirror symmetry with respect to the plane.
In the monolayer, SOC mixes $p_z$ with $p_x$ and $p_y$ orbitals of opposite
spin and, when treated perturbatively, leads to an effective
Hamiltonian\cite{Kane2005} with a spin-dependent effective
second-neighbor hopping between $p_z$ orbitals that conserves $S_z$.
In multilayer graphene this is no longer true, since electrons
in a $p_z$ orbital
in one layer can hop to the $s$ orbital of atoms in the next layer. When
SOC is added to the model, we expect that this $s-p_z$
mixing naturally leads to spin-mixing terms in the Hamiltonian, which is indeed
the case.\cite{Guinea2010}
The presence of this spin-flip channel interaction
cast a doubt on the validity of the  spin-conserving Kane-Mele model for
multilayers\cite{Prada2011,Qiao2011} and motivates our choice of the
standard\cite{Konschuh2010,Min2006,Kormanyos2013,McCann2010,Fratini2013}
four orbital tight-binding calculations.

The rest of this work is organized as follows.
In section \ref{Homogeneous} we briefly review the tight-binding
model and the procedures to determine the existence of a QSH phase
applied to the homogeneous case, studying the relation between
interlayer coupling and the topological properties of the system.
In section \ref{Heterogeneous} the same methodology is applied
to the case of a heterogeneous  structure, graphene encapsulated by a
trivial insulator, finding that topological properties can be
induced even by trivial neighboring layers.
Finally, in section \ref{conclus} we summarize our findings.

\section{Homogeneous multilayers}\label{Homogeneous}

\begin{figure}[h!]
 \centering
  \includegraphics[width=0.5\textwidth]{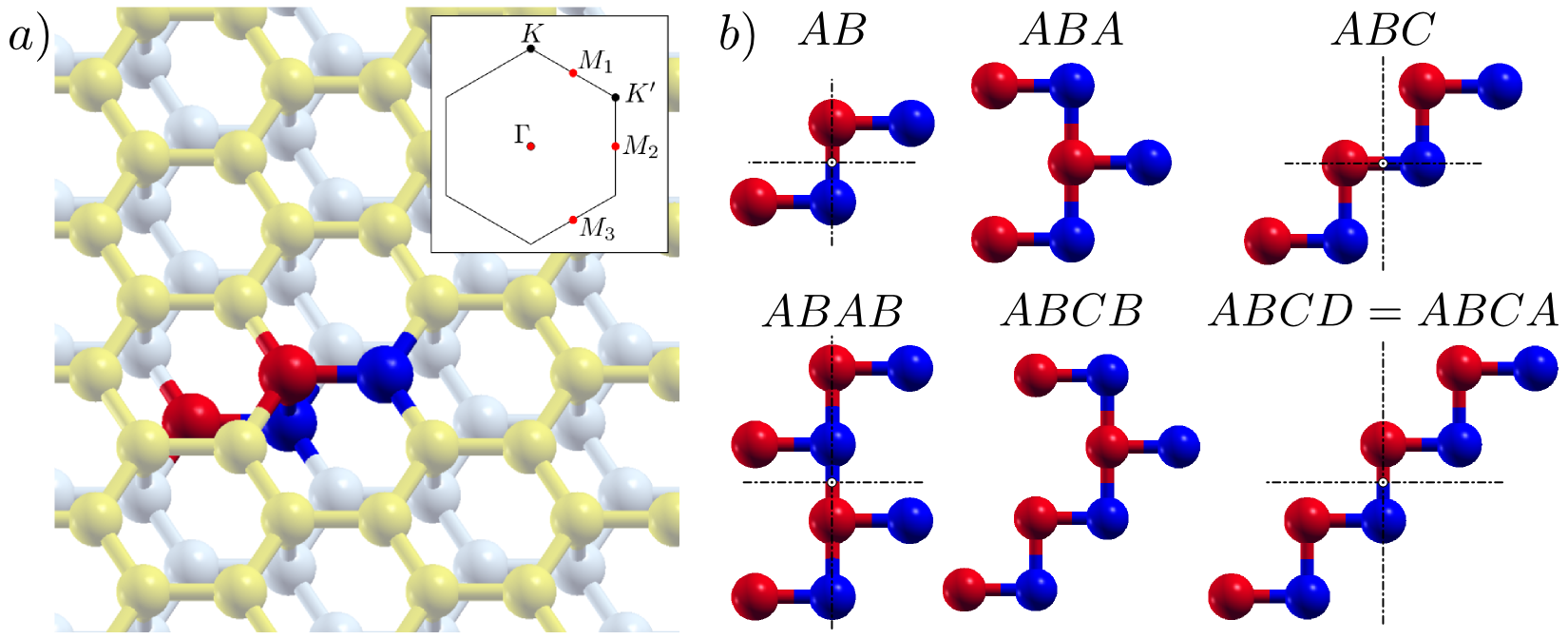}
\caption{$a)$ Crystal structure of bilayer graphene system with a
highlighted unit cell. Different colors for each layer are used to
distinguish the two layers. In the inset the first Brillouin
Zone is depicted with the high symmetry points and the Time
Reversal Invariant Momenta colored in red. In $b)$ Side view
of the unit cells for all the different stackings  studied. For the
stackings with inversion symmetry the inversion center is shown at the
crossing point of the dashed lines. For both figures red and blue denote
sublattice.}
\label{Structure}
\end{figure}

Monolayer graphene consists of a triangular lattice with two atoms per
unit cell that leads, in the reciprocal space,  to a 
 hexagonal Brillouin zone that hosts Dirac cones in its corners.
When $N$ layers are considered  the crystalline structure remains the
same, only there will be $2N$ atoms per unit cell. We shall only
use the so called Bernal stacking, shown in  Fig. (\ref{Structure}), 
 which is the ground state configuration, according to  both DFT
calculations and experimental
evidence\cite{Norimatsu2010,Charlier1994,Charlier1994a}. In Bernal
stacked materials an atom from the sublattice $B$($A$) sits on top of
an atom belonging to the other sublattice $A$($B$).
For $N=2$ there is only one way to achieve this, but for $N>2$ there
are different possible stacking orders. In figure \ref{Structure} we
show the different possibilities for $N\leq4$, with a self-evident notation.

\subsection{The Model}
We describe the multilayers with 
the following  tight-binding 
Hamiltonian:
\begin{equation}
 H = H_{ML} + \eta H_{inter} +
    \lambda\vec{L}\cdot\vec{S}
\label{Hamil}
\end{equation}
where $H_{ML}$ and $H_{inter} $ account for  the intralayer  and
interlayer  hoppings, respectively, and the last term is the
intra-atomic SOC.  Our tight-binding model  is based
on four atomic orbitals,  $s$, $p_{x}$, $p_{y}$ and $p_{z}$.
Both the intralayer and interlayer  hoppings are described within the
Slater-Koster formalism\cite{Slater1954}. The intralayer
hopping parameters are taken from 
 Ref \cite{Gosalbez-Martinez2011}.      In order to study the effect
of interlayer coupling, the interlayer terms are scaled by a
dimensionless  parameter $\eta$.   When $\eta=1$, the ratio   between
interlayer and intralayer $V_{pp\pi}$ in graphene is taken
as\cite{Katsnelson2012} 0.13.  Unless otherwise stated, in all
our calculations we have $\eta=1$. 
Within this model, the dimension of the Hilbert space for the
minimal unit cell of the crystal with $N$ layers
is $4\times2\times2\times N = 16N$ (4 orbitals per atom, 2 atoms
per layer,  plus the two possible spin orientations).

Without SOC, this model reproduces the very well known band structure
of graphene ($N=1$) and multilayer graphene $N>1$, that portraits
these systems as zero-gap semiconductors.   Within this model,   SOC
is known to open a gap in the monolayer\cite{Min2006}  as well as
in the bilayer \cite{Konschuh2012,Guinea2010,Cortijo2010}.  In the case of the
monolayer graphene the gap is known to be topological.    Within this
model, the computed value of the gap $1.46\mu $eV when we take  a
realistic value of the atomic spin orbit coupling,  $\lambda=10$meV.  This
gap is much smaller than the ones obtained with accurate density
functional theory (DFT) calculations, in the range
of 30$\mu$eV\cite{Konschuh2011a}. 
The reason for the discrepancy turns out to be 
that 
 the mayor contribution to the SOC gap at the Dirac point comes
from the coupling to the higher energy $d$ bands\cite{Konschuh2011a,Konschuh2012}.
The later is a simple consequence of the fact that SOC opens a gap in
second order in the coupling in the Dirac points when projected over
the $p$ band. In comparison, SOC acts as first order when considering
channels involving the $d$ band.
Nevertheless, interlayer hopping may open a first order spin flipping
channel in the $p$ manifold, becoming of the same order
as the intrinsic spin conserving $d$-level contribution.
These last processes would be the ones
missing in the multilayer Kane-Mele model, and should be
added for completeness.
In our case, for the sake of simplicity, 
we will focus on the spin flipping channel, and use a four orbital
tight-binding model considering $\lambda$ as a free parameter.

Future work shall focus on
the effect of the d-levels in multilayer
graphene, which will not be addressed here.

\begin{figure*}
\begin{minipage}{.9\linewidth}
\begin{center}
 \includegraphics[width=1.0\textwidth]{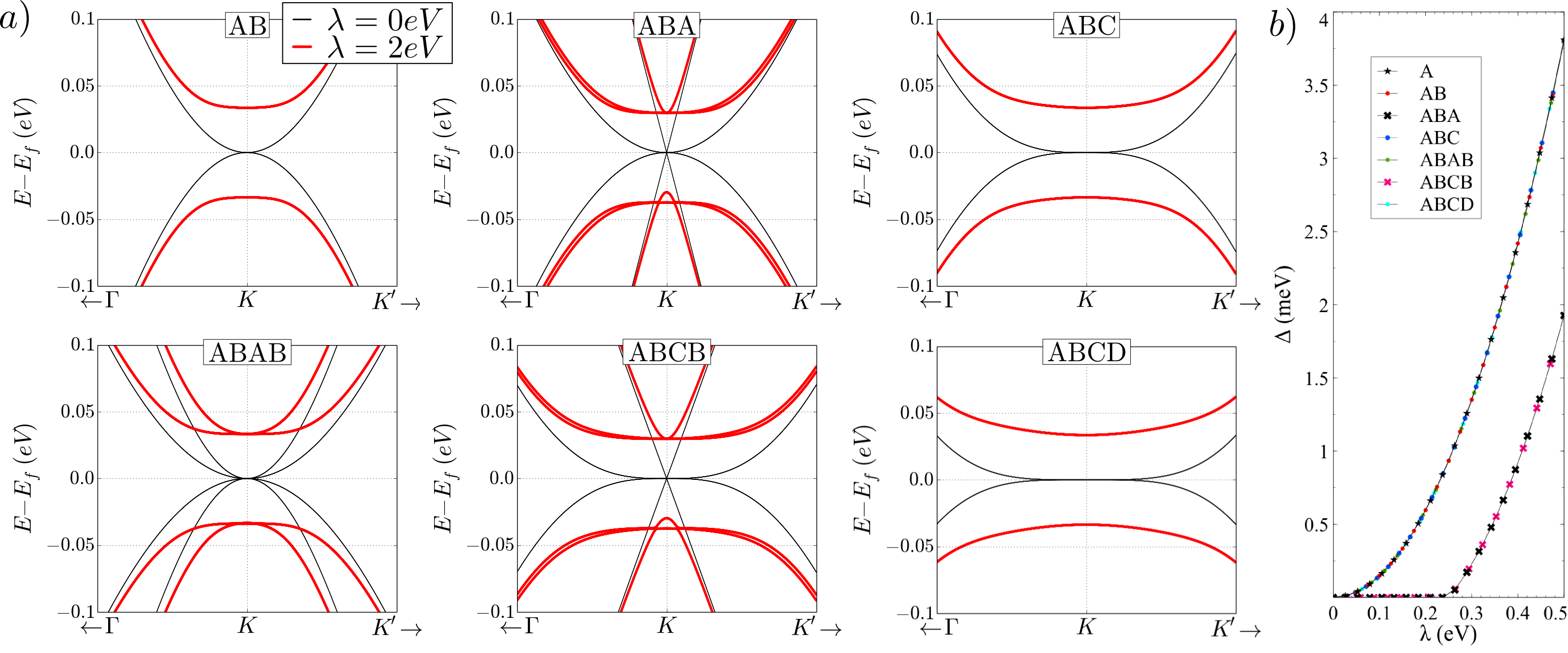}
\end{center}
\end{minipage}
\caption{ In $a)$ the band structure close to the $K$ point is shown for all the possible stackings of multilayer graphene with $N = {2,3,4}$. Only when $\lambda \neq 0$ (red line) a gap is opened at the Dirac points.
Note that for $ABA$ and $ABCB$ stackings there are linear bands when $\lambda=0$ that when the SOC is switched on cause a smaller gap than in the other cases.
In $b)$ the dependence of the gap with the SOC $\lambda$ is shown. The anomalous behavior for the $ABA$ and $ABCB$ stackings is just due to the linear bands mentioned before.}
\label{Bilayer}
\end{figure*}

The effect of SOC on the band structure of the multilayers can be summarized in the following points: 
\begin{enumerate}
\item SOC  opens up a gap for all the $N$ stacked layers
considered, reproducing the existing results\cite{Guinea2010} for
the case of $N=2$.
Notice that in the case of $ABA$ and $ABCB$ stackings,
the system remains gapless until a critical value of $\lambda$.
This peculiarity is related to the non uniform evolution of the SO splitting
of the linear and non-linear bands  as shown in figure \ref{Bilayer}.

\item  The scaling of the gap with $\lambda$ is very similar for
monolayer and $N=2,3,4$ multilayers as it is shown in
Figure \ref{Bilayer}. Therefore, it is expected that within this model,
the gap opened by
the intrinsic SOC might be as small
in multilayers as in monolayers.

\item The magnitude of the band-gap is insensitive to the interlayer
coupling. This result is somewhat surprising, since together with
atomic SOC the interlayer coupling opens a spin-flip
channel, otherwise missing in the monolayer case. In particular, switching
on the interlayer coupling does not close the SOC gap of the
monolayer as shown in Figure \ref{Bilayer}. As a consequence,  the
ground state of two decoupled ($\eta=0$) monolayers can be
adiabatically connected to the ground state of the bilayer ($\eta=1$).
\end{enumerate}

The last observation leads
to the following result:
odd $N$ stacked graphene will be
quantum spin Hall Insulators, whereas  even $N$ will not. 
More precisely,
for a system of $N$ decoupled monolayers the $Z_{2}$ invariant is:
\begin{equation}
 Z_{2}(N) = \left[Z_{2}(1)\right]^{N}
 \label{trivial}
\end{equation}
Since the gap opened by $\lambda$ remains unaffected when switching on the interlayer coupling $\eta$, the value of $Z_2$ for graphene-like multilayers is also given by equation \eqref{trivial}.  In the following we verify equation (\ref{trivial}) using two different strategies. In the
case of inversion symmetric structures, we compute the $Z_2$ invariant.
In all cases, we compute the edge states and check whether they
fill the gap, or else.  Independently on how
the topological character is obtained,
eq. \eqref{trivial} holds in all the cases.

\subsection{Calculation of the $Z_{2}$ invariant.}
Using the method developed by Fu and Kane in 2007\cite{Fu2007} for systems with
inversion symmetry it is possible to determine easily its topological character
(the $Z_{2}$ invariant) by calculating the parity of the occupied Bloch wave
functions at the time reversal invariant momenta (TRIMs).
\begin{equation}
\delta_{i} = \displaystyle\prod^{N}_{m=1}\xi_{2m}(\Gamma_{i}) \quad;\quad (-1)^{\nu} = \prod_{i}\delta_{i}
\end{equation}
where $\xi_{2m}$ is the parity eigenvalue of the $2m^{th}$ occupied state at the TRIM $\Gamma_{i}=\{\Gamma,M_{1},M_{2},M_{3}\}$. Using this method the topological character of a system will be determined just by the quantity $(-1)^{\nu}$, resulting that $(-1)^{\nu}=+1$ means trivial topology and $(-1)^{\nu}=-1$ means non trivial topology.
The calculation for the systems with inversion symmetry yields the following results:
\begin{center}
 \begin{tabular}{ c | c c c c c}
              &  A  &  AB & ABC & ABAB & ABCD \\\hline
     $M_{1}$  & $+$ & $+$ & $+$ &  $+$ &  $+$  \\
     $M_{2}$  & $+$ & $+$ & $+$ &  $+$ &  $+$  \\
     $M_{3}$  & $+$ & $+$ & $+$ &  $+$ &  $+$  \\
     $\Gamma$ & $-$ & $+$ & $-$ &  $+$ &  $+$  \\\hline
 $(-1)^{\nu}$ & $-$ & $+$ & $-$ &  $+$ &  $+$
 \end{tabular}
\label{parity}
\end{center}

This guarantees that $A$ and $ABC$  crystals are topological but the bilayers and tetralayers (with inversion symmetry) are not. This method cannot be applied to systems without inversion symmetry,  which are addressed in the next section using a different approach. 

\subsection{Edge states}
\begin{figure}[hbt]
 \centering
  \includegraphics[width=0.5\textwidth]{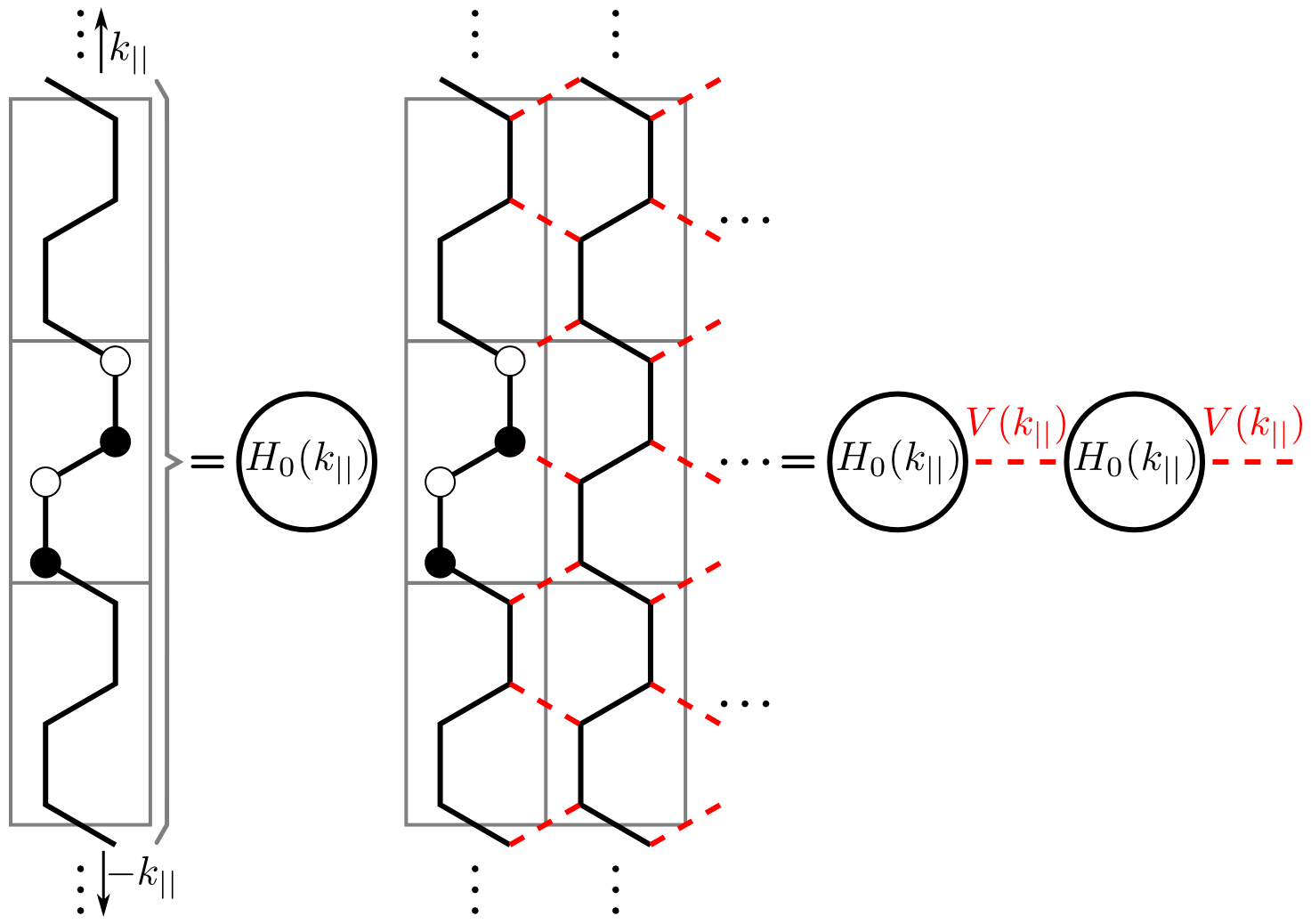}
\vspace{-15pt}
\caption{Scheme of the mapping between a semi-infinite crystal and a semi-infinite chain. The coupling between each linear chain (with $k_{||}$ well defined) is introduced by means of a self energy $\Sigma_{R}$.}
\label{Mapping}
\end{figure}

To confirm equation \eqref{trivial} even for systems without inversion symmetry we look for the presence of gapless edge states.
We consider armchair-terminated  semi-infinite crystals.  Using translation invariance along the direction parallel to the edge, we block-diagonalize  the Hamiltonian of the semi-infinite 2D  crystal in terms of a collection of $k_{||}$ dependent semi-infinite 1D Hamiltonians, as indicated in figure (\ref{Mapping}).   The 1D Hamiltonian describes unit cells with $4N$ atoms, where $N$ stands for the number of graphene layers.  The intra-cell terms are denoted by $H_0(k_{||})$ and the inter-cell hoppings by $V(k_{||})$.

\begin{figure*}
\begin{minipage}{.85\linewidth}
\begin{center}
 \includegraphics[width=1.0\textwidth]{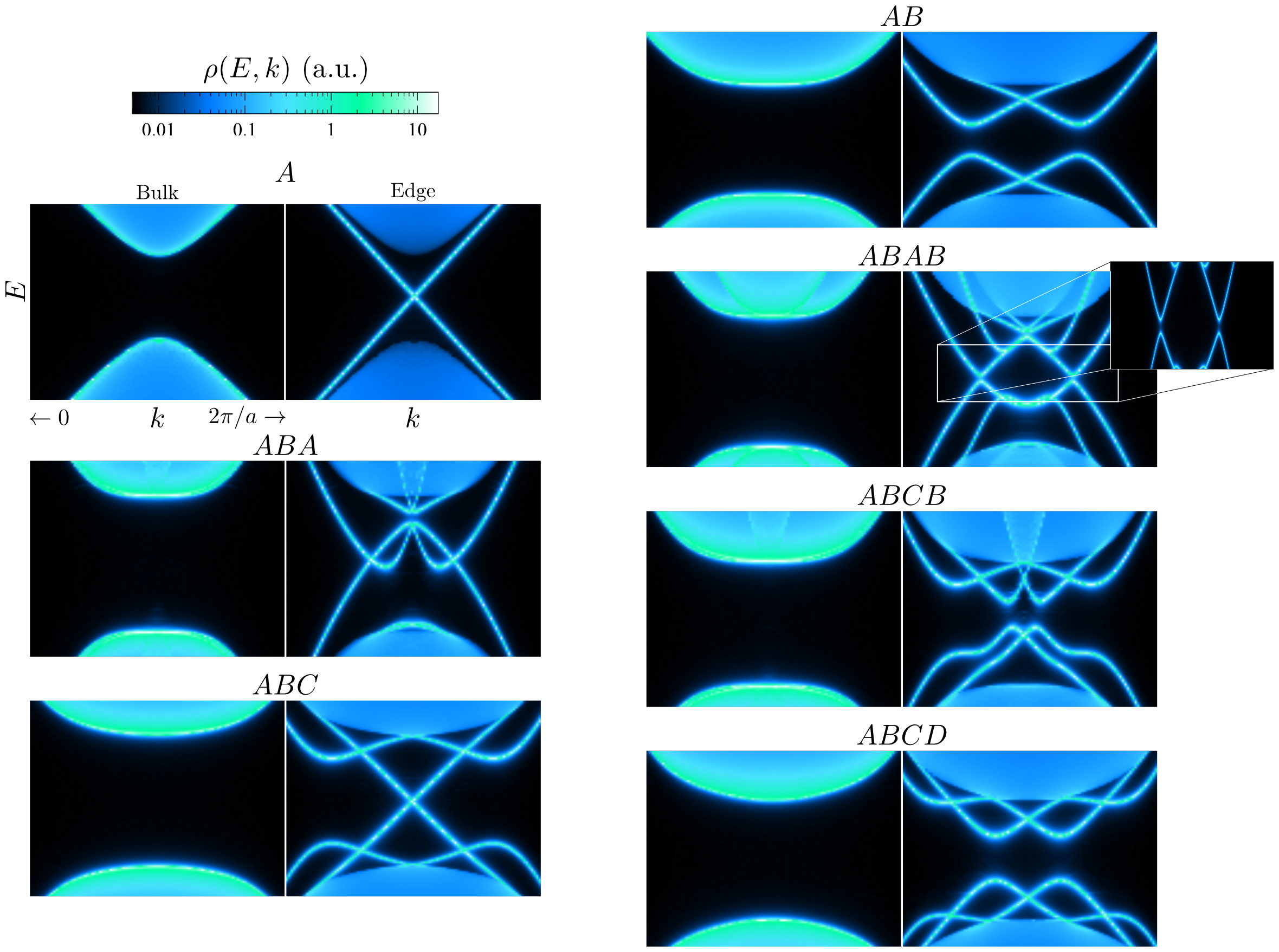}
\end{center}
\end{minipage}
\caption{ For each structure bulk and edge density of
states (left and right panel respectively).
Gapless edge states appear only when an odd number of layers is
considered independently of the stacking used.}
\label{DOSdyson}
\end{figure*}

 The surface Green function of this  block
tridiagonal semi-infinite matrix can be  written as: 
\begin{equation}
G^{edge}(E,k_{||}) =\left[E+i\epsilon-H_0(k_{||})-\Sigma_{R}(k_{||})-\Sigma_{H}(k_{||})\right]^{-1}
\end{equation}
where $\Sigma_{R}(k_{||})$ is a self-energy that accounts for the coupling to the semi-infinite crystal,
$\Sigma_{H}(k_{||})$ is the self-energy due
to its interaction with the $H$ atoms included to get
rid of the dangling bonds and $\epsilon$ a small analytic continuation.

The self-energy $\Sigma_{R}$ can be calculated employing a recursive Green's function method that leads to the following coupled equations
\begin{eqnarray}
\nonumber \Sigma_{R}(E,k_{||})&=& V_{R}(k_{||})g_{R}(E,k_{||})V^{\dagger}_{R}(k_{||}) 
\\
g_{R}(E,k_{||}) &=& \left[E-H_0(k_{||}) -\Sigma_{R}(E,k_{||})\right]^{-1}
\end{eqnarray}

The $\Sigma_{H}(k_{||})$ is calculated just as an additional iteration to the self-consistent calculation with the appropriate value for the hoppings $C-H$.

For a given $k_{||}$ we compute the density of states using
\begin{equation}
 \rho(E,k_{||}) = -\frac{1}{\pi} Im[G^{edge}(E,k_{||})]
\end{equation}
Using a similar  approach we can also obtain the bulk density of states  calculating the bulk Green function by recursion.

In figure \ref{DOSdyson} we show the density of states for both bulk and edge for all the stackings as a contour plot in the $k_{||},E$ plane. For each stacking the left panel shows the bulk density of states, which are gaped for all the stackings and the right panel shows the edge states.  The calculations are done for a rather large value of $\lambda=2$eV. The first thing to  notice is that, for such large values of $\lambda$, all the structures have edge states. However, only in the case of odd $N$, shown in the left column, the in-gap states are gapless.  This is a necessary condition in order to have a QSHI.  In contrast,  all systems with even $N$ have edge states with a gap. Thereby, they are definitely not in the QSH phase, validating 
 equation \eqref{trivial}. Therefore, we conclude that odd  $N$ graphene stacks are QSHI and even $N$ are trivial insulators. In all cases, the gap opened by SOC is quite small.

\section{Heterogeneous multilayers}\label{Heterogeneous}
In the previous section we have seen that for homogeneous multilayers the gap opened by SOC has the same magnitude than for the monolayer. Thereby, homogeneous multilayers of graphene  would not improve the prospects for observation of the QSH phase compared to the monolayer. We thus explore the case of a heterogeneous multilayer. This is motivated in part by recent experiments \cite{Avsar2014} that seem to indicate an enhancement of the SOC interaction in graphene due to proximity to WS$_2$,  a trivial semiconductor with quite large SOC and no inversion symmetry. There has also been plenty of work studying the enhancement of SOC interaction in graphene due to proximity to heavy metals\cite{Zhang2014a}.  However, it would be much more interesting if  graphene could be driven into a QSH phase by proximity to an insulator, so that the only conducting channels would be only at the edges of graphene. 

Density functional calculations show\cite{Kaloni2014} that a topological band-gap opens in graphene on top of both WS$_2$ and WSe$_2$, two widely studied two dimensional transition metal dichalcogenides (TMD).  The magnitude of this gap is in the range of a few meV, i.e.,  two or  three orders of magnitude larger than  the intrinsic SOC gap.  

\begin{figure}[hbt]
 \centering
  \includegraphics[width=0.45\textwidth]{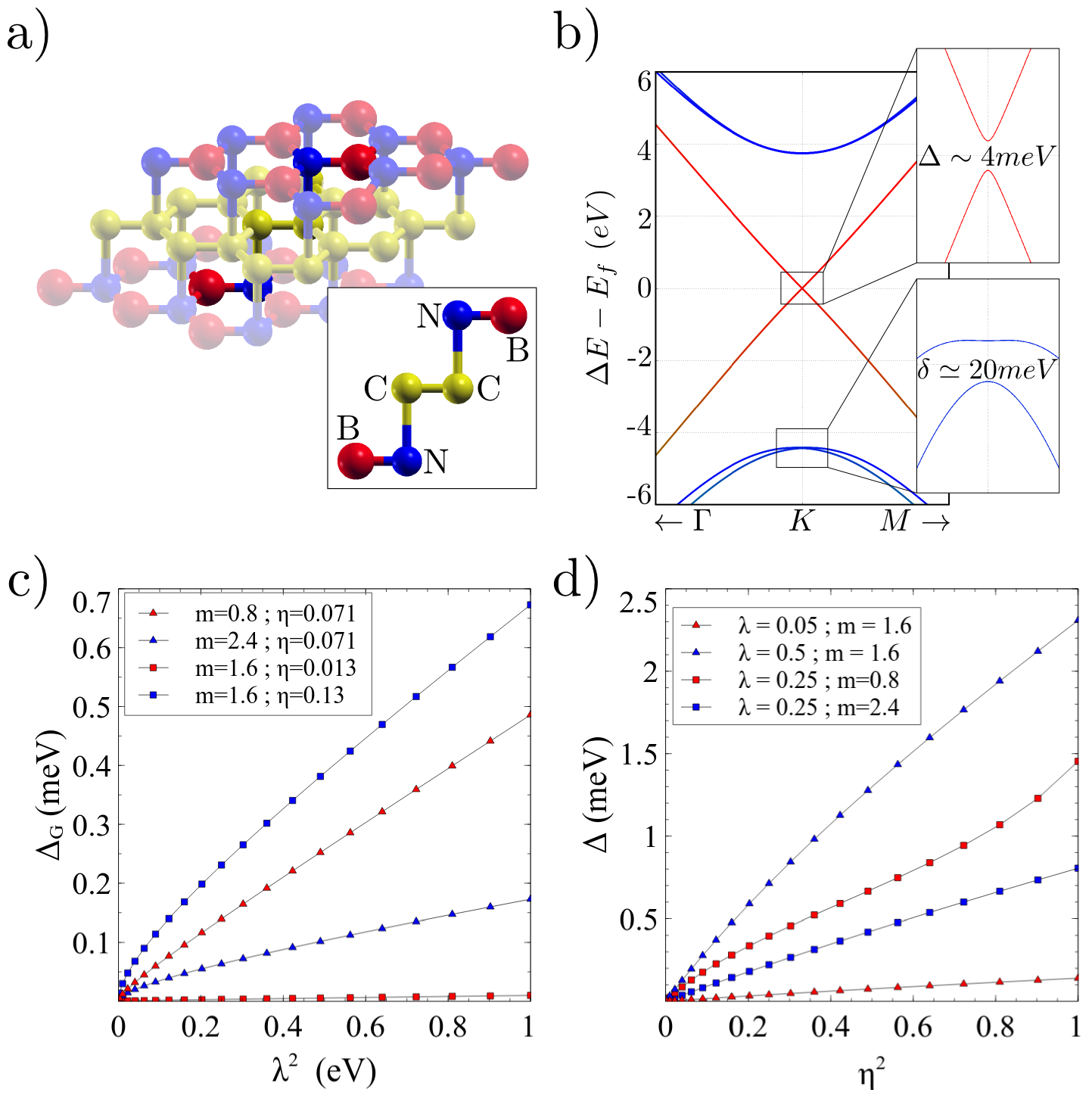}
\caption{Panel $a)$ shows the structure of the heterostructure considered.
Panels (b,c,d) show the dependence of the induced gap in
graphene due to the proximity of the encapsulating layers. In panel $b)$ it can be seen that the gap is proportional to $\lambda^{2}$ and this estimation gets better as the gap of insulating layers gets bigger. Panel $b)$ it can
shows how the interlayer coupling $\eta$ produces the
expected effect, for small interlayer coupling
the induced gap is small
but it grows quickly as $\eta$ increases. Panel $c)$ shows the dependence
of the induced gap with the sublattice imbalance}
\label{induced}
\end{figure}

Here we propose a toy model to understand the opening of a non-trivial gap due to proximity to a trivial insulator with strong spin orbit coupling.
For that matter, we take graphene encapsulated between two monolayers of a trivial semiconductor with strong SOC and broken inversion symmetry.   Specifically, the structure of  these adjacent monolayers is that of a BN-like crystal (see figure \ref{induced}(a)).   The choice of the stacking is such that, globally, the structure has inversion symmetry. Otherwise,  a trivial band gap would be opened by proximity\cite{Giovannetti2007}.

The BN-like crystal is described with the same interatomic  Slater-Koster parameters than graphene,  but very different on-site parameters. In particular we assume a large SOC $\lambda$ and a staggered potential $\pm m$  that breaks inversion symmetry of the top and bottom layers.
Since we are interested in the proximity effect, we turn off the atomic SOC of the graphene layer.
As in the case of the homogeneous multilayers, the interlayer coupling is characterized by the dimensionless parameter $\eta$.   In this case we impose
zero SOC for the graphene layer, in order to study the proximity effect.   For $\eta=0$ the bands of this system would be the superposition of those of the top and bottom insulators, with gap $2m$, and the bands of graphene, whose Dirac cones would lie inside the gap.  Broadly speaking, this picture remains the same as the interlayer coupling is turned on.    Interestingly, a non-trivial gap $\Delta$  opens in the Dirac cones only when $\eta\neq 0$ and $\lambda\neq 0$.  We have verified that this gap satisfies the  scaling 
\begin{equation}
\Delta \propto \frac{\lambda \eta^2}{m^{2}}
\label{wcoupling}
\end{equation}
in the limit of small $\lambda$, $\eta$ and $m^{-1}$.   This results implies that graphene can borrow SOC from a neighbor trivial insulator  layer via interlayer coupling. Using the method of the TRIM  we have verified that this insulator has
$Z_2=(-1)^{\nu}=-1$, and is therefore topologically non-trivial. 

The magnitude of the proximity effect away from the weak coupling limit of eq. \eqref{wcoupling} is shown in figures \ref{induced}.  We study the dependence of the proximity gap $\Delta$ as a function of both the SOC $\lambda$ and the interlayer coupling $\eta$ for two values of the encapsulating layer  staggered potential $m$.  It is apparent that, taking $m=2.0eV$ (a trivial gap $\sim1.5eV$) and $\lambda\simeq 0.25$eV,  values in line with those of 2D TMD,  the proximity gap is in the order of 1meV, similar to the DFT results. Therefore, our model provides a reasonable justification of the DFT computations, which are certainly more complete.  

Our  toy model does not capture some probably important features of real heterogeneous multilayers.  For instance, 
the interlayer interaction  could break inversion symmetry which is expected to open a trivial gap.  In addition,   
the geometry of our encapsulating  layers was  chosen to minimize the size of the unit cell, rather than to describe a real material.  In general, the coupling of graphene to other 2D crystals will imply a new length scale, given by the size of the new unit cell. In this setup, the inversion symmetry breaking could average out.

\section{Conclusions}\label{conclus}
We have studied the quantum spin Hall phase in multilayer graphene and in graphene encapsulated by a trivial semiconductor. In the case of multilayer graphene we find that only the stacks with an odd number of layers are quantum spin Hall insulators. However, the size of the gap is the same than for a monolayer, and thereby, most likely too small to be detected experimentally.
In contrast, we propose a toy model for  graphene encapsulated between two semiconducting layers with strong SOC and a trivial gap. Our model shows that  a non-trivial gap can be opened in graphene whose magnitude is controlled by the atomic spin orbit coupling of the adjacent layers. Our model provides a qualitative understanding of recent DFT calculations\cite{Zhang2014a} as well as recent experimental work\cite{Avsar2014} and shows a promising route to observe the quantum spin Hall phase in graphene.

\section{Acknowledgments}

JFR acknowledges financial supported by MEC-
Spain (FIS2010-21883-C02-01) and Generalitat Valenciana (ACOMP/2010/070), Prometeo. This work has
been financially supported in part by FEDER funds.
We acknowledge financial support by Marie-Curie-ITN
607904-SPINOGRAPH. J. L. Lado and N. Garcia thank
the hospitality of the Departamento de Fisica Aplicada at
the Universidad de Alicante.

\end{document}